\documentclass[pra,aps,floats,letterpaper,floatfix,footinbib,preprintnumbers,twocolumn]{revtex4}
\usepackage{amsmath}
\usepackage{hyperref}
\usepackage{color}
\usepackage[mathenv]{}
\usepackage{graphicx}
\usepackage{ulem}
\usepackage{natbib}
\begin{document}

\title{Non-invasive Measurements of Cavity Parameters by Use of Squeezed Vacuum}

\author{Eugeniy E. Mikhailov}
\author{Keisuke Goda}
\author{Nergis Mavalvala}

\affiliation{LIGO Laboratory, Massachusetts Institute of Technology, Cambridge, MA 02139, USA}

\begin{abstract}
We propose and experimentally demonstrate a method for
non-invasive measurements of cavity parameters by injection of
squeezed vacuum into an optical cavity. The principle behind this
technique is the destruction of the correlation between upper and
lower quantum sidebands with respect to the carrier frequency when
the squeezed field is incident on the cavity. This method is
especially useful for ultrahigh $Q$ cavities, such as whispering
gallery mode (WGM) cavities, in which absorption and scattering by
light-induced nonlinear processes inhibit precise measurements of
the cavity parameters. We show that the linewidth of a test cavity
is measured to be $\gamma = 844\pm40$~kHz, which agrees with the
classically measured linewidth of the cavity within the
uncertainty ($\gamma=856\pm34$~kHz).

\end{abstract}

%\pacs{}
\definecolor{purple}{rgb}{0.6,0,1}
\preprint{\large \color{purple}{LIGO-P060006-01-R}}
%\date{\today}
\maketitle

\section{Introduction}
\label{sect:intro}

High Q cavities such as whispering gallery mode (WGM) cavities
have recently demonstrated quality factors ($Q$) as high as
$2\times 10^{10}$ and have shown the potential to reach even
higher Q values
\cite{ilchenko2004prl,vahala2003nature,savchenkov2004pra}.
However, there are difficulties in measurement of the linewidth
and Q of such high Q cavities. While in theory, the Q factor could
be as high as $10^{12}$ and is limited only by Rayleigh scattering
\cite{gorodetsky2000josab}, in practice, it is limited by other
losses in the cavity. They include absorption and scattering
losses due to impurities in the cavity material, and light-induced
losses due to nonlinear processes. Due to the extremely small mode
volume and high Q-factor of the cavity, the cavity build-up
intensity is extremely high, even in the case of an input with
small power (as small as several mW). Such a high resonator
intensity leads to very efficient nonlinear processes inside WGM
cavities, such as Raman scattering, second harmonic generation,
and four-wave mixing \cite{savchenkov2004prl}. Whereas this is
beneficial in many applications, it causes additional losses in
the cavity and thus makes the Q factor measurement unreliable (at
least, making it power-dependent) \cite{savchenkov2006private}.

Squeezed states of vacuum or light have been used in many
applications such as improvement in interferometric
\cite{caves1981prd,mckenzie2002prl,xiao1987prl,grangier1987prl}
and absorption \cite{marin1997optcomm} measurements, {for}
quantum teleportation \cite{furusawa1998science} {and} quantum
cryptography \cite{bennett1992jcrypto}, and {for} quantum
imaging \cite{kolobov1999rmp}. However, to the best of our
knowledge, no experiment for measurements of cavity parameters by
use of squeezing
has yet been reported. In this paper we propose and
demonstrate an alternative method of measuring Q factors by use of
a squeezed vacuum field which is equivalent to a field with
correlated quantum sidebands \cite{caves1985pra,schumaker1985pra}.
This technique is advantageous over traditional optical methods in
that it utilizes the injection of squeezed vacuum into a test
cavity not to excite any nonlinear processes in the cavity. When
the input field is detuned from the cavity resonance frequency, it
transmits only the upper or lower quantum sidebands within the
cavity linewidth while reflecting the counterparts (associated
upper or lower sidebands) and all the other sidebands. The
linewidth of the cavity can then be measured by observing the
destruction of the correlation between the upper and lower quantum
sidebands with respect to the carrier frequency. We show that the
linewidth and Q factor of a test cavity using the method agrees
with those measured by traditional optical methods.

This paper is organized as follows: In Sec. \ref{sect:theory1}, we
describe the theoretical framework for the measurement method. In
Sec. \ref{sect:theory2}, we explain the validity of the use of
squeezed vacuum as a probe for non-invasive measurements {and compare the technique to using a classical state}. In Sec.
\ref{sect:experiment}, we demonstrate the method using a test
cavity with known cavity parameters and compare the parameter
values obtained by the new method and the traditional optical
methods. The conclusions of the paper are summarized in Sec.
\ref{sect:conclusions}.

\section{Theory}
\subsection{Destruction of Quantum Sideband Correlation as a Probe for Cavity Parameter Measurements}
\label{sect:theory1}

\begin{figure}[t]
\includegraphics[angle=0, width=0.6\columnwidth]{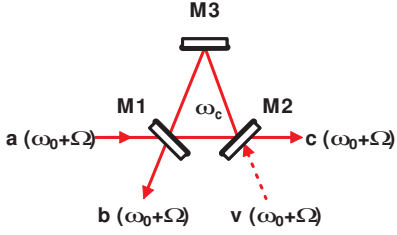}
\caption{(Color online) Schematic of a cavity under test. The cavity is composed
of three mirrors M1, M2, and M3 in triangular geometry with
reflectivities R$_1$, R$_2$, and R$_3$, respectively. $a$ is the
upper sideband of an injected field at frequency
$\omega_0+\Omega$, $b$ is the cavity-filtered reflection at the
frequency, $c$ is the transmission at the frequency, and $v$ is
the vacuum field that couples in due to losses in the cavity at
the frequency. $\omega_c$ is the cavity resonance frequency.
{The
carrier field at frequency $\omega_0$ transmits through the cavity
when $\omega_0 = \omega_c$}.} \label{cavity}
\end{figure}

Consider a squeezed vacuum field with carrier and sideband
frequencies, $\omega_0$ and $\omega_0 \pm\Omega$ respectively. As
shown in Fig.~\ref{cavity}, when the upper sideband of the
squeezed vacuum field $a(\omega_0 + \Omega)$ is injected into an
optical cavity with resonance frequency $\omega_c$ and mirror
reflectivities $R_1,R_2$, and $R_3$, the
{reflected field} $b(\omega_0 + \Omega)$ and
its adjoint $b^{\dagger}(\omega_0 - \Omega)$ are given in terms of
$a(\omega_0+\Omega)$ and its adjoint $a^{\dagger}(\omega_0 -
\Omega)$ by
\begin{eqnarray}
\label{eq:b}
\hspace{-0.35cm} b(\omega_0 + \Omega) &=& \nonumber \\
&& \hspace{-2cm} r(\omega_0 + \Omega)\,a(\omega_0 + \Omega) + l (\omega_0 + \Omega)\, v(\omega_0 + \Omega), \\
\label{eq:bdagger}
 \hspace{-0.35cm} b^{\dagger}(\omega_0 -\Omega) &=& \nonumber \\
&& \hspace{-2cm} r^*(\omega_0-\Omega)\,a^{\dagger}(\omega_0-\Omega) + l^*(\omega_0-\Omega)\, v^{\dagger}(\omega_0-\Omega),
\end{eqnarray}
where $r(\omega_0 \pm \Omega)$ is the frequency-dependent cavity reflection coefficient and $l(\omega_0 \pm \Omega)$ is the vacuum noise coupling coefficient associated with transmission and intra-cavity losses. When the cavity is not perfectly mode-matched, the reflected field contains the cavity-coupled reflection $a_c$ \cite{siegman1986university} and the promptly reflected field $a_m$ that does not couple to the cavity due to mode mismatch such that
\begin{eqnarray}
r(\omega_0 + \Omega)a(\omega_0 + \Omega) &=& \nonumber\\
&&\hspace{-3cm} r_c(\omega_0 + \Omega)a_c(\omega_0+\Omega) + r_m a_m(\omega_0+\Omega), \\
r^{*}(\omega_0 - \Omega)a^{\dagger}(\omega_0 - \Omega) &=& \nonumber\\
&&\hspace{-3cm} r_c^{*}(\omega_0 - \Omega)a_c^{\dagger}(\omega_0-\Omega) + r_m^{*} a_m^{\dagger}(\omega_0-\Omega),\\
l(\omega_0 + \Omega)v(\omega_0 + \Omega) &=& \nonumber\\
&&\hspace{-3cm} l_c(\omega_0 + \Omega)v_c(\omega_0+\Omega) + l_m v_m(\omega_0+\Omega), \\
l^{*}(\omega_0 - \Omega)v^{\dagger}(\omega_0 - \Omega) &=& \nonumber\\
&&\hspace{-3cm} l_c^{*}(\omega_0 - \Omega)v_c^{\dagger}(\omega_0-\Omega) + l_m^{*} v_m^{\dagger}(\omega_0-\Omega),
\end{eqnarray}
where $a_c$ and $a_m$ are spatially orthogonal and
\begin{eqnarray}
&&\hspace{-1.25cm}r_c(\omega_0 \pm \Omega) = r_c(\omega_d \pm \Omega)\nonumber\\
&=& \sqrt{R_1} - \frac{T_1\sqrt{R_2 R_3}e^{-i\left[\phi_c(\omega_d)\pm \phi_s(\Omega)\right]}}{1-\sqrt{R_1 R_2 R_3}e^{-i\left[\phi_c(\omega_d) \pm \phi_s(\Omega)\right]}}, \\
r_m &=& \sqrt{R_1}.
\end{eqnarray}
Here, $\omega_d$ is the detuning from the cavity resonance given by $\omega_d = \omega_0 - \omega_c$ and we have assumed that the resonance frequency of $a_m$ is far from that of $a_c$ such that the reflection coefficient $r_m$ can be treated as a frequency-independent constant at frequencies around the resonance frequency of $a_m$. The vacuum noise coupling coefficients are then given by
\begin{eqnarray}
l_c(\omega_0 \pm \Omega) &=& l_c(\omega_d \pm\Omega) = \sqrt{1 - |r_c(\omega_d \pm \Omega)|^2}, \\
l_m(\omega_0 \pm \Omega) &=& l_m(\omega_d \pm\Omega) = \sqrt{1 - r_m^2}.
\end{eqnarray}
The cavity mirror reflectivity and transmission of each mirror satisfies
\begin{eqnarray}
R_i + T_i + L_i = 1, \hspace{0.5cm} \mbox{for} \hspace{0.3cm} i = 1,2,3,
\end{eqnarray}
where L$_i$ is the loss of each mirror. The intra-cavity losses
can be absorbed into $R_3$.

Since the carrier is detuned from the cavity resonance frequency,
the reflection acquires extra frequency-dependent phase shifts at
the detuned carrier frequency and the sideband frequencies,
respectively given by
\begin{eqnarray}
\label{eq:phases}
\phi_c = \frac{p}{c}\omega_d = 2 \pi
\frac{\omega_d}{\omega_{\rm FSR}},
\hspace{0.5cm} \phi_s =
\frac{p}{c}\Omega = 2 \pi \frac{\Omega}{\omega_{\rm FSR}},
\end{eqnarray}
where $p$ and $\omega_{\rm FSR}$ are the round-trip length and
free spectral range of the cavity, and $c$ is the speed of light
in vacuum.

For simplicity, we transform into the rotating frame of the
carrier frequency $\omega_0$ in the frequency domain, such that
Eqs.~\eqref{eq:b} and~\eqref{eq:bdagger} become
\begin{eqnarray}
\label{eq:b_transformed}
b(\Omega) &=&  r_c(\omega_d + \Omega)a_c(\Omega) + r_m a_m(\Omega) \nonumber\\
&&+ l_c(\omega_d + \Omega)v(\Omega) + l_m v_m(\Omega), \\
\label{eq:bdagger_transformed}
b^{\dagger}(-\Omega) &=& r_c^{*}(\omega_d-\Omega)a_c^{\dagger}(-\Omega) + r_m^{*} a_m^{\dagger}(-\Omega)\nonumber\\
&&+ l_c^*(\omega_d-\Omega)v^{\dagger}(-\Omega) + l_m^{*} v_m^{\dagger}(-\Omega),
\end{eqnarray}
where $a_c(\Omega)$ and $a_c^{\dagger}(-\Omega)$ satisfy the commutation relations
\begin{equation}
\label{eq:commutation_relations} \left[a_c(\pm \Omega),
a_c^{\dagger}(\pm \Omega')\right] = 2\pi\delta(\Omega-\Omega^{'}),
\end{equation}
and all others vanish (similarly for $a_m(\Omega)$, $a^{\dagger}_m(-\Omega)$, $v_c(\Omega)$, $v_c^{\dagger}(-\Omega)$, $v_m(\Omega)$, and $v_m^{\dagger}(-\Omega)$). In the two-photon representation
\cite{caves1985pra,schumaker1985pra}, the amplitude and phase
quadratures of $a_c$ are defined by
\begin{eqnarray}
\label{eq:quadratre_fields}
a_1^c(\Omega) &=& a_c(\Omega) + a_c^{\dagger}(-\Omega), \\
a_2^c(\Omega) &=& -i\left[a_c(\Omega) - a_c^{\dagger}(-\Omega)\right],
\end{eqnarray}
respectively (similarly for $a_m$, $b$, $v_c$, and $v_m$). A little algebra yields
the amplitude and phase quadrature fields of the
{reflected light} in compact matrix form,
\begin{eqnarray}
\label{eq:b_compact}
\textbf{b} = \textbf{M}\textbf{a}_c + r_m\textbf{a}_m + \textbf{H}\textbf{v}_c + l_m \textbf{v}_m,
\end{eqnarray}
where we use the two-photon matrix representation
\begin{eqnarray}\hspace{-0.5cm}
\textbf{a}_c &\equiv& \left(\begin{array}{ccc}
a_1^c \\
a_2^c
\end{array}\right)
\end{eqnarray}
for the operator $a_c$ (similarly for $a_m$, $b$, $v_c$, and $v_m$),
\begin{eqnarray}
\textbf{M} = {\rm e}^{i\,\varphi_{-}}\left(\begin{array}{ccc}
\cos\varphi_{+} & -\sin\varphi_{+} \\
\sin\varphi_{+} & \cos\varphi_{+}
\end{array}\right)
\left(\begin{array}{ccc}
A_{+} & i\,A_{-} \\
-i\,A_{-} & A_{+}
\end{array}\right)
\end{eqnarray}
is a matrix representing propagation through the cavity, and
\begin{eqnarray}
\textbf{H} =
\left(\begin{array}{ccc}
l_+ & il_- \\
-il_- & l_+
\end{array}\right).
\end{eqnarray}
$\textbf{M}$ comprises an overall phase shift $\varphi_{-}$,
rotation by angle $\varphi_{+}$, and attenuation by factor
$A_{+}$. Here we have defined
\begin{eqnarray}
\varphi_{\pm} &\equiv& \frac{1}{2}\left[\mbox{arg}(r_c(\Omega)) \pm \mbox{arg}(r_c(-\Omega))\right], \\
A_{\pm} &\equiv& \frac{1}{2}\left[|r_c(\Omega)|\pm
|r_c(-\Omega)|\right],\label{eq:phiA} \\
l_{\pm} &\equiv& \frac{1}{2}\left[l_c(\omega_d+\Omega) \pm l_c(\omega_d-\Omega)\right].
\end{eqnarray}
In the case of no carrier detuning ($\omega_d =
0$), $r_c(\Omega) = r_c^*(-\Omega)$, and  $\varphi_{+}$
and A$_{-}$ vanish, giving neither quadrature angle
rotation nor asymmetrical amplitude attenuation.
In the case of cavity detunings ($\omega_d \neq 0$), nonzero
$\varphi_{+}$ gives quadrature angle rotation.

From Eq.~\eqref{eq:b_compact}, when we perform homodyne detection of the reflected field with a local oscillator (LO) field, the measured amplitude and phase
quadrature variances of the field, defined by $V_1^b = {\left\langle
b_1^2\right\rangle - \left\langle b_1\right\rangle^2}$ {and}
$V_2^b = {\left\langle b_2^2\right\rangle - \left\langle
b_2\right\rangle^2}$ {(similarly for $V_1^{a_c}$, $V_2^{a_c}$, $V_1^{a_m}$, and $V_2^{a_m}$)}, are found in terms of the mode-matched input amplitude and phase
quadrature variances $V_1^{a_c}$ and $V_2^{a_c}$ to be
\begin{eqnarray}
\label{eq:variances}
&&\hspace{-1cm}
\left(\begin{array}{cc}
V_{1}^{b}\\
V_{2}^{b}
\end{array}\right) = \nonumber\\
&&\hspace{-0.5cm}
\eta_c
\left(\begin{array}{cc}
\cos^2\varphi_{+}&\sin^2\varphi_{+}\\
\sin^2\varphi_{+}&\cos^2\varphi_{+}
\end{array}\right)
\left(\begin{array}{cc}
A_+^2&A_-^2\\
A_-^2&A_+^2
\end{array}\right)
\left(\begin{array}{cc}
V_{1}^{a_c}\\
V_{2}^{a_c}
\end{array}\right) \nonumber\\
&&\hspace{-0.5cm}+ \eta_m r_m^2
\left(\begin{array}{cc}
V_1^{a_m} \\
V_2^{a_m}
\end{array}\right)
+
\eta_c\left[1-\left(A_+^2 + A_-^2\right)\right]
\left(\begin{array}{cc}
1\\1
\end{array}\right) \nonumber\\
&&\hspace{-0.5cm}
+\eta_m (1-r_m^2)
\left(\begin{array}{cc}
1 \\
1
\end{array}\right) +
\eta_l
\left(\begin{array}{cc}
1 \\
1
\end{array}\right),
\end{eqnarray}
where $\eta_c$ and $\eta_m$ are the composite efficiencies of detection
associated with the cavity-coupled and cavity-mismatched modes respectively, $\eta_l$ is the coupling of detection losses, and $\eta_c + \eta_m + \eta_l = 1$. The detection efficiency is a product of the quantum efficiency of the photodiodes and the mode-overlap efficiency with the LO mode. Eq. \eqref{eq:variances} can be rewritten in terms of the quadrature variances of the incident field $V_{1,2}^a$ since the cavity-coupled reflection $V_{1,2}^{a_c}$ and the mode-mismatch reflection $V_{1,2}^{a_m}$ originate from the same incident field $V_{1,2}^a$, such that
\begin{eqnarray}
\left(\begin{array}{cc}
V_1^{a_c} \\
V_2^{a_c}
\end{array}\right) =
\left(\begin{array}{cc}
V_1^{a_m} \\
V_2^{a_m}
\end{array}\right) =
\left(\begin{array}{cc}
V_1^{a} \\
V_2^{a}
\end{array}\right),
\end{eqnarray}
and therefore,
\begin{eqnarray}
\label{eq:measuredvariances}
&&\hspace{-0.4cm}
\left(\begin{array}{cc}
V_{1}^{b}\\
V_{2}^{b}
\end{array}\right) = \nonumber\\
&&\hspace{-0.4cm}
\left[\eta_c
\left(\begin{array}{cc}
\cos^2\varphi_{+}&\sin^2\varphi_{+}\\
\sin^2\varphi_{+}&\cos^2\varphi_{+}
\end{array}\right)
\left(\begin{array}{cc}
A_+^2&A_-^2\\
A_-^2&A_+^2
\end{array}\right)
+ \eta_m r_m^2
\right]
\left(\begin{array}{cc}
V_1^{a} \\
V_2^{a}
\end{array}\right) \nonumber\\
&&\hspace{-0.4cm}+
\left[1-\eta_c\left(A_+^2 + A_-^2\right) - \eta_mr_m^2\right]
\left(\begin{array}{cc}
1\\1
\end{array}\right).
\end{eqnarray}
Note that if the input field is in a vacuum or coherent state such that $V_1^a = V_2^a = 1$, then $V_1^b = V_2^b = 1$, as expected, and no cavity information is contained in the output state $b$.

If the carrier frequency is detuned downward from the cavity resonance
frequency, the cavity transmits only the upper sidebands within
the cavity linewidth and replaces them by vacuum at those
frequencies while reflecting the associated lower sidebands and
all the other sidebands. Hence, the cavity-coupled reflected field
is composed of the uncorrelated sidebands within the linewidth and
the reflected correlated sidebands outside of it. The consequence
is the destruction of the correlation within the linewidth between
the upper and lower quantum sidebands. This is analogous to the
destruction of the correlation between electro-optically modulated
coherent sidebands in pairs, in which the beat between the carrier
and the upper or lower sideband can be measured only when either
sideband is absorbed into the cavity, reflecting the carrier and
other sideband. The beat could not be observed if all the fields
were reflected. Similar measurements could be done with the
transmission of the squeezed vacuum field through the cavity.
However, the signal-to-noise ratio would not be as good as in the
reflection method because the background of the transmission
signal is shot noise.

It is convenient to define the test cavity linewidth $\gamma$, the
quality factor $Q$, and the finesse ${\mathcal F}$, as
\begin{eqnarray}
\label{eq:gamma}
\gamma &=& \frac{2}{\pi}\,\omega_{\rm FSR}\sin^{-1}\left[\frac{1-\sqrt{R_1R_2R_3}}{2(R_1R_2R_3)^{1/4}}\right] \nonumber\\
&\simeq& \frac{1-\sqrt{R_1R_2R_3}}{\pi(R_1 R_2
R_3)^{1/4}}\,\omega_{\rm FSR},
\end{eqnarray}
\begin{eqnarray}
\label{eq:Qfactor} Q = \frac{\omega_0}{\gamma}\,,
\end{eqnarray}
and
\begin{eqnarray}
\label{eq:finesse} {\mathcal F} =
\frac{\pi(R_1R_2R_3)^{1/4}}{1-\sqrt{R_1R_2R_3}} \simeq
\frac{\omega_{\rm FSR}}{\gamma}\,,
\end{eqnarray}
respectively. The approximations made in Eqs. \eqref{eq:gamma} and
\eqref{eq:finesse} are valid for high Q cavities. $R_1$, $R_1 R_2
R_3$, $\omega_d$, and $\omega_{\rm FSR}$ will be treated as free
fitting parameters. We also assume the input mirror is lossless
such that $T_1=1-R_1$.

\subsection{Squeezed/Anti-squeezed Vacuum vs. Classically Noisy Light}
\label{sect:theory2} Since we are interested in having as little
light (at the carrier frequency) as possible in the test cavity,
it is instructive to calculate the average photon number in the
field we use. The average photon number in squeezed light with
squeeze factor $r$ and squeeze angle $\theta$ is given by
\cite{scully1997cambridge}
\begin{eqnarray}
\label{eq:photon_number}
\left\langle N \right\rangle &=& \left\langle a^{\dagger}a\right\rangle \nonumber\\
&=& |\alpha|^2(\cosh^2 r + \sinh^2 r) - (\alpha^*)^2 e^{i\theta}\sinh r \cosh r \nonumber \\
&& - \alpha^2 e^{-i\theta}\sinh r \cosh r + \sinh^2 r,
\end{eqnarray}
where $\alpha$ is the coherent amplitude of the light. As the
number of coherent photons becomes zero ($\alpha\rightarrow 0$),
resulting in squeezed vacuum, Eq.~\eqref{eq:photon_number} becomes
\begin{eqnarray}
\label{eq:photon_number_squeezed}
\left\langle N \right\rangle = \left\langle a^{\dagger}a\right\rangle = \sinh^2 r.
\end{eqnarray}
This is the average photon number in squeezed vacuum generated by
squeezing. Note that if the field is unsqueezed ($ r = 0$),
$\left\langle N \right\rangle = 0$. For a squeeze factor of 1.5
corresponding to the squeezed or anti-squeezed level of $-13$~dB
which is the current experimental limit
\cite{lam1999job,aoki2005arxiv}, $\left\langle N \right\rangle =
4.53$. Therefore, it is fair to say that the optical influence of
ideal squeezed vacuum on cavities is negligible.

\begin{figure}
\includegraphics[angle=0, width=1.0\columnwidth]{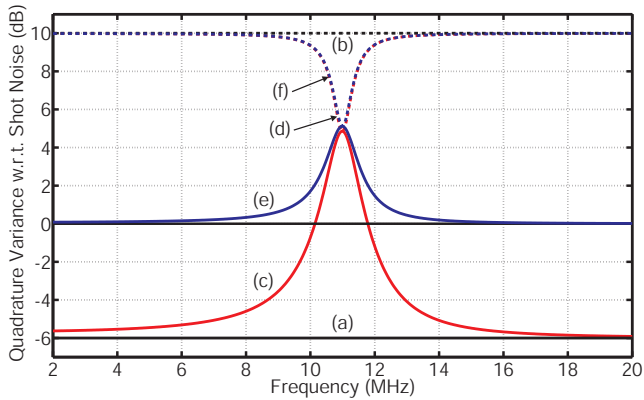}
\caption{(Color online) Comparison of signal contrast between squeezed and classical fields injected into an impedance-matched cavity. The quadrature variances $V_1^b$ and $V_2^b$ are shown as solid and dashed curves, respectively, for different input states. (a) and (b) show the (impure) input state with $V_1^a = -6$ dB and $V_2^b = 10$ dB, \textit{in the absence of the cavity}. (c) and (d) show the cavity-coupled response to the squeezed and anti-squeezed vacuum injection, respectively. (e) and (f) show the
cavity-coupled response to injection of a classically noisy state with $V_1^a = 0$ dB, $V_2^a = 10$ dB. Comparing (e) and (c), we note that squeezing improves the signal contrast, but the classical noise and the anti-squeezed quadrature behave almost identically [cf. (d) and (f)].}
\label{cl_vs_sq}
\end{figure}

Similarly, it is instructive to compare this technique to using a
classical state. For simplicity,
assuming that the quadrature variance in both quadratures is
frequency-independent, we consider the case in which the lower
sideband is fully transmitted through an impedance-matched cavity
and the upper sideband {is} fully reflected {at the} input
mirror such that $r_c(-\Omega)=0$ and $r_c(\Omega)=1$ at $\Omega =
\omega_d$, respectively,  {which gives} $A_{+}
=A_{-}=1/2$ from Eq.~\eqref{eq:phiA}. Thus, the amplitude and
phase quadrature variances of the reflected
field are found to be
\begin{equation}
V_1^b(\omega_d)=V_2^b(\omega_d)=\frac{1}{4}\left(V_1^a+V_2^a\right)+\frac{1}{2}.
\end{equation}
In the absence of  coherent light, the signal contrast can
be defined as the quadrature variance at detuning frequency
$\omega_d$ compared to the cavity-uncoupled quadrature variance at
off-resonance frequencies ($|\Omega-\omega_d| \gg \gamma$), in
which case $V_1^b = V_1^a$ and $V_2^b = V_2^a$, and the signal
contrasts at the two orthogonal quadratures are respectively given
by
\begin{eqnarray}
S_1(\omega_d)&=\frac{V_1^b(\omega_d)}{V_1^a}
&= \frac{\frac{1}{4}(V_1^a+V_2^a) + \frac{1}{2}}{V_1^a},
\\
S_2(\omega_d)&=\frac{V_2^a}{V_2^b(\omega_d)}
&= \frac{V_2^a}{\frac{1}{4}(V_1^a+V_2^a)+ \frac{1}{2}}.
\end{eqnarray}
In the limiting case of {$V_2^a \gg V_1^a$ and $V_2^a \gg 1$},  we obtain
\begin{eqnarray}
S_1(\omega_d) &\simeq& \frac{V_2^a}{4\,V_1^a},
\\
S_2(\omega_d) &\simeq& 4.
\end{eqnarray}

\noindent We see that $S_2$ has about the same limiting level as
in the classical case, while $S_1$ grows if $V_1^a$ gets smaller.
Classically, $V_1^a \geq 1$ (the shot noise limit), but using
squeezed vacuum we can obtain $V_1^a < 1$, or improved signal
contrast for a measurement in the squeezed quadrature. This is
illustrated in Fig.~\ref{cl_vs_sq}, where we compare the signal
contrast for measurement of the cavity linewidth using a classical
field with the signal contrast for squeezed field injection. The
cavity-coupled responses of the classical and anti-squeezed
quadrature variances behave almost identically in the case of the
impedance-matched cavity, whereas squeezing improves the signal
contrast of the measurement.

\subsection{Fundamental Limit on Measurement Uncertainty}
It is important to note that even in the absence of  technical noise,
quadrature variance measurements are intrinsically  contaminated by quantum noise itself.
The standard deviation of the quadrature variances is given by~\cite{mckenzie2005job}
\begin{eqnarray}
\Delta V_j^b = \sqrt{2} V_j^b \hspace{1cm} \mbox{for} \hspace{0.2cm} j=1,2.
\end{eqnarray}
Thus, the noise of the measurement is proportional to the measured
value itself, and many averages can be performed to achieve smaller
uncertainty levels.

This is different from the classical case where the parameters of a
cavity are measured by measuring the transmission of a probe optical
field incident on the cavity as a function of cavity detuning. In
this case, the measurements are fundamentally limited by shot noise:
the number of measured photons (N) has uncertainty proportional to
$\sqrt{N}$. Therefore, the signal-to-noise ratio grows as the number
of the transmitted photons increases.

\section{Experiment}
\label{sect:experiment}

\begin{figure}
\includegraphics[angle=0, width=1.0\columnwidth]{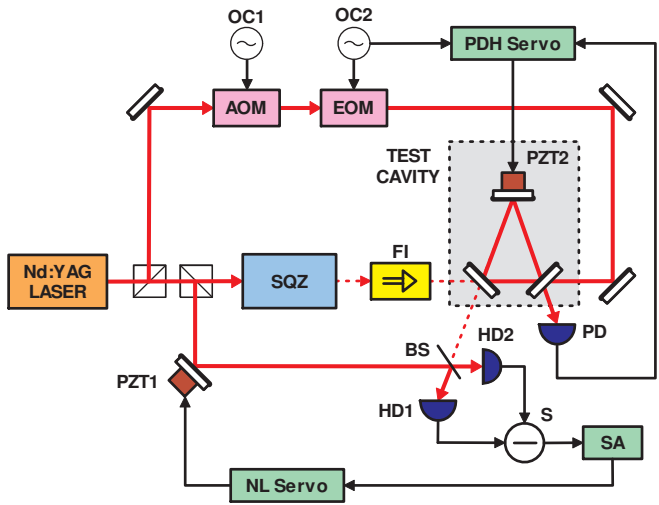}
\caption{(Color online) Schematic of the experiment. SQZ: squeezed vacuum
generator, FI: Faraday isolator, AOM: acousto-optic modulator,
EOM: electro-optic modulator, OC1 and OC2: oscillators, PZT1 and
PZT2: piezo-electric transducers, PD: photo-detector, HD1 and HD2:
homodyne photo-detectors, BS: 50/50 beamsplitter, S: substractor,
SA: spectrum analyzer, NL Servo: noise-locking servo, PDH Servo:
PDH-locking servo. The oscillators (OC1, OC2) are driven at
$11.0\pm0.1$ MHz and $13.3\pm0.1$ MHz respectively. The squeezed
vacuum generator is composed of an optical parametric oscillator
(OPO) and a second harmonic generator (SHG) that pumps the OPO.
The cavity length is locked to the laser frequency by the
PDH-locking servo and PZT (PZT2). The homodyne angle is locked by
the noise-locking servo and PZT (PZT1).} \label{setup}
\end{figure}

The experiment is schematically shown in Fig.~\ref{setup}. The
Nd:YAG laser (Lightwave Model 126) gives an output of cw 700 mW at
1064 nm, which is injected into the squeezed vacuum generator
(squeezer). The squeezer is composed of a second harmonic
generator (SHG) and an optical parametric oscillator (OPO), both
using 5$\%$ MgO:LiNbO$_3$ nonlinear crystals placed within optical
cavities (hemilith for the SHG and monolith for the OPO) in the
Type I phase-matching configuration. The SHG pumped by the Nd:YAG
laser generates 250 mW at 532 nm, which then pumps the OPO below
threshold with a vacuum seed. The resultant field generated by the
OPO is a squeezed vacuum field with a squeezing bandwidth of 66.2
MHz defined by the OPO cavity linewidth. A sub-carrier field,
frequency-shifted by an acousto-optic modulator (AOM) to a
frequency that is coincident with the cavity TEM$_{01}$ mode, is
injected into the other end of the OPO cavity. The cavity is thus
locked to the TEM$_{01}$ mode, offset by 220 MHz from the carrier
frequency, using the Pound-Drever-Hall (PDH) locking technique
\cite{drever1983applphysb}. The frequency-shift is necessary to
ensure that no cavity transmitted light at the fundamental
frequency is injected into the OPO cavity since it acts as a seed
and degrades broadband squeezing due to the imperfect isolation of
the Faraday isolator \cite{mckenzie2004prl,bowen2002job}. This is
especially important for high Q cavities with linewidths as narrow
as kHz because low-frequency squeezing is difficult to achieve.

\begin{figure}
\includegraphics[angle=0,width=1.0\columnwidth]{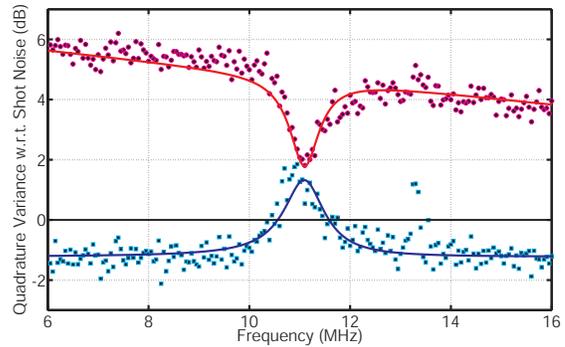}
\caption{(Color online) Measured squeezed and anti-squeezed quadrature variances
with respect to shot noise (dots) and fits to the data points
(curves) using Eq. \eqref{eq:measuredvariances}. The resolution bandwidth
of the spectrum analyzer is 100 kHz. The data are exponentially
averaged 100 times. The apparent peak at 13.3 MHz is due to the
coupling of the EOM modulation at the frequency for the
PDH-locking technique. The overall decrease in the squeezing and
anti-squeezing levels with frequency is due to the OPO cavity
linewidth. With the optically measured FSR, the linewidth is found
from the fits to be $\gamma = 844\pm40$  kHz. } \label{linewidth}
\end{figure}

The squeezed vacuum is injected into a triangular test cavity with
the FSR of 713 MHz and FWHM of $\gamma=856\pm34$ kHz, both
measured by traditional methods using light. The frequency shift,
of the subcarrier is $231\pm0.1$ MHz so that the carrier frequency
is detuned from the TEM$_{00}$ mode by $11.0\pm0.1$ MHz. As a
result of this frequency shift, only the upper sidebands are
within the cavity linewidth, destroying the correlation between
the upper and lower sidebands and, therefore, destroying the
squeezing or anti-squeezing. This cavity-coupled squeezed vacuum
reflection is measured by balanced homodyne detection, where the
field to be measured interferes with a local oscillator (LO) field
and is detected by two (nearly) identical photodetectors. The
difference of the two photodetector signals is sent to an HP4195A
spectrum analyzer (SA) to measure the noise variance of the
squeezed or anti-squeezed quadrature. The results are shown in
Fig.~\ref{linewidth}. The experimental data are exponentially
averaged 100 times. The resolution bandwidth of the spectrum
analyzer is 100 kHz. Since the squeezed vacuum does not carry any
coherent amplitude, the noise-locking technique
\cite{mckenzie2005job} is employed to lock the homodyne angle to
either the squeezed or anti-squeezed quadrature at 2 MHz.

Before fitting the experimental data points, the homodyne efficiencies $\epsilon_{h_c}$ and $\epsilon_{h_m}$, and the quantum efficiency of the photo-detectors $\epsilon_{QE}$ need to be taken into account.
The sum of the homodyne efficiencies and the quantum efficiency were independently measured to be $90 \%$ and $85 \%$ respectively. The sum of the efficiencies $\eta_c +\eta_m$ in Eq. \eqref{eq:measuredvariances} is given by $\eta_c+\eta_m = (\epsilon_{h_c}+\epsilon_{h_m})\epsilon_{QE}$. We ignore $\epsilon_{h_m}$ since the cavity mode-matching efficiency is $82 \%$ and hence $\epsilon_{h_m} \ll \epsilon_{h_c}$, which yields $\eta_l \simeq 1 - \eta_c$. Moreover, we have assumed that the
input mirror M$_1$ is lossless. This assumption is valid since it
is a single-pass loss and does not influence the linewidth of the
cavity. We then fit Eq. \eqref{eq:measuredvariances} to the
{measured} data points with free parameters $R_1$,$R_2\,R_3$,
and $\omega_d$; {both the data and the fits are} shown in
Fig.~\ref{linewidth}. The resulting fitting values are
$\sqrt{R_1\,R_2\,R_3} = 0.99628 \pm 0.00016, \sqrt{R_1} =
0.99783\pm0.00005$, and $\omega_d/(2\pi) = 11.098 \pm 0.017$ MHz.
Therefore, the FWHM linewidth of the cavity is found to be $\gamma
= 844\pm40$ kHz, which agrees with the classically measured
linewidth of the cavity within the uncertainty ($\gamma=856\pm34$
kHz). We note that $\omega_{\rm FSR}$ can be determined from the
fit, but here we have used the optically measured value to
estimate the linewidth. This is valid because any loss in the
cavity does not change the FSR.

\section{Conclusion}
\label{sect:conclusions} We have proposed and experimentally
demonstrated a method for non-invasive measurements of optical
cavity parameters by use of squeezed vacuum. The technique has the
advantage over traditional optical methods that the injection of a
squeezed vacuum field as a probe for cavity parameters does not
excite any nonlinear processes in cavities, and is, therefore,
useful for ultrahigh Q cavities such as whispering gallery mode
(WGM) cavities. We have shown that when a squeezed vacuum field is
injected into a detuned cavity, the linewidth and $Q$ factor of a
test cavity can be determined by measuring the destruction of
upper and lower quantum sidebands with respect to the carrier
frequency. The linewidth of a test cavity is measured to be
$\gamma = 844 \pm 40$ kHz, which agrees with the classically
measured linewidth of the cavity within the uncertainty
($\gamma=856 \pm 34$ kHz). We have also show that the use of
squeezed fields leads to better signal contrast, as expected.

\section{Acknowledgments}
We would like to thank our colleagues at the LIGO Laboratory,
especially Thomas Corbitt and Christopher Wipf, and Stan Whitcomb
for his valuable comments on the manuscript. We gratefully
acknowledge support from National Science Foundation Grant Nos.
PHY-0107417 and PHY-0457264.

%%%%%%%%%%%%%%%%%%%%%%%

%%%%%%%%%%%%%%%%%%%%%%%

\end{document}